\pdfoutput=1
\documentclass[pdflatex,sn-mathphys-ay]{sn-jnl}
\usepackage{amsmath,amssymb,amsthm}
\usepackage{graphicx}
\usepackage{booktabs}
\usepackage{algorithm}
\usepackage{algpseudocode}
\usepackage{enumitem}
\usepackage{multirow}

\newtheorem{proposition}{Proposition}

\DeclareMathOperator*{\argmin}{arg\,min}
\newcommand{\E}{\mathbb{E}}

\newcommand{\R}{\mathbb{R}}

\renewcommand{\paragraph}[1]{\par\addvspace{4pt plus 2pt}\noindent\textit{#1}\hspace{0.4em}\ignorespaces}
\makeatletter\let\table\tableorg\let\endtable\endtableorg\makeatother
\begin{document}
\title[Group recovery and level-free flagging]{Group recovery after trimming, and level-free flagging, in robust clusterwise regression}
\author*[1]{\fnm{Samir} \sur{Orujov}}\email{sorujov@ada.edu.az}
\affil*[1]{\orgdiv{School of Business}, \orgname{ADA University}, \orgaddress{\street{61 Ahmadbay Aghaoglu Street}, \city{Baku}, \postcode{AZ1008}, \country{Azerbaijan}}}
\abstract{Trimming methods for robust clusterwise regression discard a fixed fraction of the data. Too low
a level breaks the fit; too generous a level can trim away a small group. We first propose
a group-recovery step that can follow any trimming or flagging method: it searches the discarded
units for a line, tests whether those near it form a peak rather than a band, and restores the
line as a group when the likelihood of Gaussian groups plus uniform noise improves by a margin
like that of the Bayesian information criterion. In simulations it repaired the failures of a
generous TCLUST-REG level with unequal groups, but not with three or four groups. The second
proposal, ESF (exact-subsample flagging), is a flagging procedure without a trimming level: it solves the clusterwise least-squares
problem exactly on small subsamples, flags units far from the best fit, and draws later
subsamples from the rest. Two constants stand in for the level: a subsample size, set from a
lower bound on the smallest group proportion, and a cap on the flagged set. It is meant for
data about whose contamination nothing is known: TCLUST-REG at the fixed level 0.30, followed by
reweighting and the recovery step, was as accurate as ESF on average up to a fifth of outliers, and
higher levels were more accurate beyond. The
flagged fraction estimates the contamination only when the errors are close to Gaussian. On taxi
fares with known tariffs, ESF found both in every sample.}
\maketitle
\section{Introduction}
\label{sec:intro}

In clusterwise linear regression each observation follows one of $K$ linear models, and the
model that generated it is not observed. For units $(x_i,y_i)$, $i=1,\dots,n$, with $x_i\in\R^{d}$
including an intercept,
\begin{equation}
y_i=x_i^{\top}\theta^{0}_{z_i}+\sigma_{z_i}e_i,\qquad z_i\in\{1,\dots,K\},
\label{eq:model}
\end{equation}
where $z_i$ is the unobserved label of unit $i$, $\theta^{0}_k\in\R^{d}$ and $\sigma_k>0$ are
the coefficients and noise scale of group $k$, and the $e_i$ are independent standard errors.
The model goes back to \citet{quandt1972switching} and \citet{spath1979algorithm}. It is
usually fitted by least squares over partitions or by maximum likelihood for a mixture of
regressions \citep{desarbo1988maximum,deveaux1989mixtures}.

Both fits break down when a fraction of the data does not follow any of the $K$ models: a few
gross outliers in the response can pull a fitted line far from every group, or be fitted as a
group of their own. The standard remedy is trimming. TCLUST-REG
\citep{garciaescudero2010robust} discards the fraction $\alpha$ of units that fit worst and
fits the model to the rest; the trimmed likelihood estimator (TLE) of \citet{neykov2007robust} and the
trimmed cluster-weighted model \citep{garciaescudero2017robust} do the same within other
likelihoods. Each needs the trimming level $\alpha$, which the analyst rarely knows. If $\alpha$
is smaller than the fraction of outliers, the fit breaks down; if it is much larger than that
fraction, genuine units are discarded, and a small group can be trimmed away altogether.

Several ways around this choice have been proposed. One trims at a deliberately high level
and then reinstates the units that are well fitted, a reweighting step proposed for robust
clustering by \citet{dotto2018reweighting}; it needs a level that is high enough. Another
monitors the fit over a grid of levels and leaves the choice to the analyst
\citep{torti2019assessing,riani2008fitting}, or scans the number of groups and the level
jointly by trimmed likelihood curves \citep{garciaescudero2011exploring,cerioli2018finding}.
A third adapts the trimming within each group \citep{chang2020componentwise}, and a fourth
replaces trimming by bounded loss functions \citep{bai2012robust}. Mixtures with a mean-shift
parameter for each unit \citep{yu2017new}, with a uniform noise component
\citep{banfield1993model}, or with contaminated normal errors
\citep{punzo2017robust,mazza2020mixtures} model the outliers instead of trimming them. Multiple-model
fitting from random minimal samples, from RANSAC
\citep{fischler1981random} to J-linkage \citep{toldo2008robust}, T-linkage
\citep{magri2014tlinkage} and preference analysis \citep{magri2015robust}, treats a related
problem with the noise scale as a user-set threshold, and does not ask whether what it
discards is a missed model.

This paper makes two proposals. The first concerns what a robust fit throws away. When one
group is small, a trimming method at a generous level can trim it away with the outliers, and a
fit that flags units by their residuals can spend two lines on the large group and flag the small
one; our simulations show both happening to a group of $20\%$ of the data even without outliers.
The \emph{group-recovery step} (Algorithm~\ref{alg:recovery}) searches the trimmed or flagged
units for a line around which many of them are concentrated, and keeps the line if the
likelihood of Gaussian groups plus uniform noise prefers the fit that contains it. The step
needs only a fit and its flagged set, so it can follow TCLUST-REG with reweighting as well as
the level-free procedure introduced next; it is what makes a generous level safe with
unequal groups. Trimmed likelihood curves ask the same question by a joint scan; the step
asks it locally and post hoc (Section~\ref{sec:recovery}).

The second proposal is a flagging procedure that needs no level, the exact-subsample flagging
procedure, ESF (Algorithm~\ref{alg:adaptive}). A small random subsample that contains no
outlier, fitted by least squares over all its partitions, gives a clean fit of the model, and
that fit identifies the outliers of the whole sample as the units far from every fitted line.
ESF therefore draws many small subsamples, fits each one exactly, scores each fit by a bounded
loss on the units not yet flagged, and flags the units that lie far from the best fit; later
subsamples are drawn only from units not yet flagged. The fit ends with the reweighting step of
robust regression \citep{rousseeuw1987robust}, and outlying covariates are flagged by a
separate screen. A unit is flagged when its residual exceeds a fixed multiple of a robust
scale; this cut-off takes the place of the trimming level, and it does not depend on how many
outliers there are. Two constants stand in for the level: the subsample size $m$, set from a
lower bound on the smallest group proportion, and the cap of $n/3$ on the flagged set beyond
which the recovery step is not attempted; Section~\ref{sec:constants} gives a default rule for
each. Subsampling to find a clean subset is the idea behind least median of squares and least
trimmed squares \citep{rousseeuw1984least,rousseeuw2006computing}, behind robust linear
grouping \citep{garciaescudero2009robust} and behind the random starts of TCLUST-REG. Here
each subsample contains several units of every group and is solved to global optimality, so
a clean subsample yields the clusterwise least-squares fit of that subsample and not a fit to
$d$ points. The exact fit is the problem that the operations research literature solves on
whole samples \citep{bertsimas2007classification,carbonneau2011globally,carbonneau2012extensions,carbonneau2014globally,fois2026clusterwise,bagirov2013nonsmooth,joki2020clusterwise,park2017algorithms},
without treating outliers. Here it is applied about a hundred times per data set to subsamples
of a dozen units, where a plain depth-first branch and bound is one to two orders of magnitude
faster than a mixed-integer programme. The accuracy of ESF does not rest on the exactness:
with ten starts of hard alternation in place of the exact fit of each subsample, ESF alone was
within $0.01$ of the exact fit in most cells, ahead by $0.025$ in one cell and behind by
$0.02$ and $0.04$ with four groups (Supplement, Table~S14). The exact fit removes instead the
dependence of a subsample's fit on its starting point.

The place of ESF among the trimming methods should be stated at once. In the simulation
study, TCLUST-REG at the fixed level $0.30$ with reweighting and the recovery step matched the
mean accuracy of ESF with up to $20\%$ of outliers and had no cell below $0.85$, where ESF had
two; the level $0.25$ with equal group weights did better still; and the levels $0.35$ and
$0.40$ beat ESF beyond $25\%$ of outliers (Table~\ref{tab:level}). The level $0.35$ came
closest to serving both ranges; up to $20\%$ of outliers it fell behind ESF only with three and
four groups, and at $35\%$ it fell behind in four designs with two groups. ESF is
therefore not the more accurate method; its case is that it needs no level and holds up with
three and four groups, where every fixed level with estimated group weights loses accuracy.

The contributions are the following.
\begin{enumerate}[label=(\roman*),leftmargin=2.2em,itemsep=2pt]
\item The group-recovery step, which can follow any trimming or flagging method
(Section~\ref{sec:recovery}).
\item ESF (Section~\ref{sec:method}), with a comparison of exact solvers for the small
clusterwise least-squares problems and an ablation of the exact fit (Section~\ref{sec:solvers}).
\item Three partial results (Section~\ref{sec:theory}): the exact probability that a stage of
the sequential draw contains a clean subsample, the large-sample limit of the final reweighting
step and of the fraction it flags, and the value of the two operations of the recovery step in
the population.
\item A simulation study on $7{,}750$ data sets (Section~\ref{sec:simulation}).
After TCLUST-REG at a generous level with reweighting, the recovery step repaired nearly all
failures with unequal or small groups but not those with three or four groups. ESF followed by
the recovery step was within $0.03$ of the better of the levels $0.25$ and $0.40$, chosen after
the run, in $83$ of $89$ cells with up to a fifth of outliers, and was the less robust choice
beyond a quarter.
\item Three data sets (Section~\ref{sec:realdata}): the fishery and tone perception data, and taxi
fares from an airport, where the tariff of every trip is recorded and so the true group of every
unit is known.
\end{enumerate}
No method in the study is best everywhere; Section~\ref{sec:discussion} says which to use
when. The constants of ESF and of the recovery step were the same in every experiment.

\section{ESF and the recovery step}
\label{sec:method}

We write ESF for the flagging procedure of Algorithm~\ref{alg:adaptive}. Unless stated
otherwise, ESF is followed by the recovery step of Algorithm~\ref{alg:recovery}, and
\emph{ESF alone} denotes Algorithm~\ref{alg:adaptive} without it. The recovery step
(Section~\ref{sec:recovery}) is described after the fit whose flagged set it examines.

\subsection{Exact fits on small subsamples}

For a set $S$ of units, the least-squares clusterwise fit of $S$ solves
\begin{equation}
\min_{z\in\{1,\dots,K\}^{S}}\ \min_{\theta\in\R^{Kd}}\ \sum_{i\in S}\bigl(y_i-x_i^{\top}\theta_{z_i}\bigr)^{2}.
\label{eq:lsfit}
\end{equation}
For fixed labels the inner problem separates into $K$ ordinary least-squares fits, so the
difficulty lies in the labels alone. On the whole sample the problem is hard: whether $n$ points
in the plane can be covered by $K$ lines, that is whether \eqref{eq:lsfit} with one covariate has
value zero, is NP-complete when $K$ is part of the input \citep{megiddo1982complexity}. On a
subsample of a few dozen units it can be solved exactly by a depth-first branch and bound over
label vectors, which prunes a partial labelling as soon as the sum of the residual sums of
squares of its groups exceeds the best value found (Supplement, Section~S7). The cost does not depend on $n$, but it grows quickly with $K$ and $m$ (Supplement,
Table~S30), which limits ESF to a small number of groups.

For fixed coefficients the best label of each unit is its nearest line, so the exact fit of a
subsample is a $K$-means fit with lines in place of centres \citep{pollard1981strong}; here
and below a \emph{line} is the graph of $x\mapsto x^{\top}\theta_k$, a hyperplane when there
is more than one covariate. The same rule, $h_{\theta}(u)=\argmin_k(y-x^{\top}\theta_k)^{2}$ for a unit $u=(x,y)$, with ties
broken by the smallest index, extends the fit to the whole sample. We call the fit of one
subsample a \emph{replicate}; it is \emph{admissible} if each of its $K$ groups contains at
least $d+1$ units, and inadmissible replicates are discarded and redrawn, which forces
$m\pi_{\min}\gtrsim d+1$ for the subsample size $m$ and the smallest group proportion
$\pi_{\min}$.

\subsection{Cost of the exact fit}
\label{sec:solvers}

ESF needs about a hundred solves of \eqref{eq:lsfit} on a dozen units per data set, so what
matters is the cost of one small solve. On the same subsamples we compared our depth-first
search, which also prunes partial labellings that cannot give every group $d+1$ units, with
repetitive branch and bound \citep{brusco2006repetitive} and with the mixed logical-quadratic
programme of \citet{carbonneau2011globally} solved by Gurobi~13.0 \citep{gurobi2026} and by
SCIP~10.0 \citep{maher2016pyscipopt} (Supplement, Section~S4 and Tables~S26 and~S30). For
$m\le20$ the depth-first search was $30$ to $180$ times faster than Gurobi, and further still
than SCIP, which solves a different, big-$M$ formulation of the programme. The gap closes as
$m$ grows, and the programme overtakes the search from about $m=40$ with two groups; the
repetitive bound made the search slower at every size tried. Exactness matters for accuracy less
than for reliability (Section~\ref{sec:intro}; Supplement, Table~S14): with one start of hard
alternation in place of the exact fit, ESF alone lost up to $0.11$ with four groups, and with
elemental fits as in RANSAC up to $0.25$.

\subsection{Sequential flagging}
\label{sec:algorithm}

Algorithm~\ref{alg:adaptive} states ESF. It runs in $L$ stages, each drawing $B/L$ admissible
replicates from the units not yet flagged; after each stage all replicates so far are scored by
a bounded loss on those units, and the units far from the best-scoring replicate are flagged.
The flags are recomputed at every stage, so a unit flagged at one stage can be reinstated at
the next.

\begin{algorithm}[tbp]
\caption{ESF: clusterwise regression from exactly solved subsamples, with sequential flagging}
\label{alg:adaptive}
\begin{algorithmic}[1]
\Require data $(x_i,y_i)_{i\le n}$; number of groups $K$; subsample size $m$; replicates $B$;
stages $L$; weight $\lambda$; cut-offs $c$, $c_{\mathrm b}$
\State $\mathcal F\leftarrow$ units whose covariates are outlying (robust distance above the
$0.975$ quantile of $\chi^{2}_{p}$, $p$ the number of covariates)
\For{$\ell=1,\dots,L$}
  \State draw $B/L$ admissible replicates, on subsamples of size $m$ drawn uniformly from
    $\{1,\dots,n\}\setminus\mathcal F$, each solved exactly
  \State $r_{bi}\leftarrow\min_k(y_i-x_i^{\top}\hat\theta^{(b)}_k)^{2}$, the squared residual
    of unit $i$ from its nearest line of replicate $b$, for every replicate $b$ so far and
    every unit $i$
  \State $s^{2}\leftarrow\min_b\operatorname{med}_{i\notin\mathcal F}r_{bi}/0.4549$, the median
    taken over the unflagged units
    \Comment{$0.4549$ is the median of a $\chi^{2}_1$ variable}
  \State $F_b\leftarrow\sum_{i\notin\mathcal F}\min\{r_{bi},c_{\mathrm b}^{2}s^{2}\}$;\quad
    $w_b\leftarrow\exp\{-\lambda(F_b-\min_{b'}F_{b'})/[(n-|\mathcal F|)\,s^{2}]\}$
  \State reference $\leftarrow$ replicate $b^{\ast}=\argmin_bF_b$
  \State $\mathcal F\leftarrow\{i:\ \text{$i$ is far from the reference in the response or in
    the covariates}\}$ (\emph{Flagging} below)
\EndFor
\State labels $\tilde z\leftarrow$ plurality of the nearest-line labels of all replicates,
  weighted by $w$
\State refit each group on its unflagged units; concentration steps keeping $n-|\mathcal F|$
  units
\State reweighting: recompute the flags at the current fit and refit on the unflagged units
\State \Return coefficients $\hat\theta$, labels $h_{\hat\theta}(u_i)$ for every unit, the flags
  $\mathcal F$ recomputed at $\hat\theta$, and $\hat\alpha=|\mathcal F|/n$
\end{algorithmic}
\end{algorithm}

\emph{Flagging.} A unit is far from a fit $\theta$ in the response if its absolute residual
from its nearest line exceeds $c$ times the scale of the group of that line. That scale is
$1.4826$ times the median absolute residual of the units assigned to the line (or of all
units, when fewer than ten are); with a scale per group, a group with small noise does not
hide outliers and a group with large noise is not trimmed. A unit is far in the covariates if its robust distance from the other
units of its group, computed with the minimum covariance determinant \citep{rousseeuw1999fast},
exceeds the $0.975$ quantile of $\chi^{2}_p$; the test is applied to groups with at least
$\max(10,5p)$ units, and at the start, before any fit, to all units as one group. It is the
covariate trimming of TCLUST-REG \citep{garciaescudero2010robust} with a fixed quantile in
place of a trimming level.

\emph{Scoring.} The bounded loss caps each squared residual at $c_{\mathrm b}^{2}s^{2}$, where
$s$ is a scale common to all replicates, so a replicate that fits the clean units well has a low
score whatever the outliers do. The scale is the smallest, over the replicates, of the median
squared residual over all unflagged units, scaled to estimate the error variance of a
replicate that fits the clean majority (Supplement, Section~S7). The criterion has a weakness
that more search does not cure. With a small group, the
bounded loss with the best replicate's scale can prefer two nearly parallel lines on the large
group, whose units they fit more tightly, to one line on each group, and a wider search finds
such fits more often: with $B=500$ replicates in place of $100$, the cell means of the
small-group designs at $30\%$ of outliers fell by $0.04$ to $0.07$ (Supplement, Table~S29),
and the recovery step then has to undo what the scoring chose. A less biased scale did not
help (Supplement, Section~S6.2).

\emph{Labels and final fit.} The labels of the final fit are the weighted plurality of the
labels that the replicates give to each unit, after alignment with a pilot replicate that
serves only for alignment (Supplement, Section~S7). The concentration steps are those of least
trimmed squares \citep{rousseeuw2006computing} and TCLUST-REG, with the number of retained
units set by the flags rather than by a level, and the last step is the reweighting step of
robust regression \citep{rousseeuw1987robust,gervini2002class}; every unit then receives the
label of its nearest fitted line.

\subsection{Is the flagged set a group?}
\label{sec:recovery}

\begin{figure}[tbp]
\centering
\includegraphics[width=0.9\textwidth]{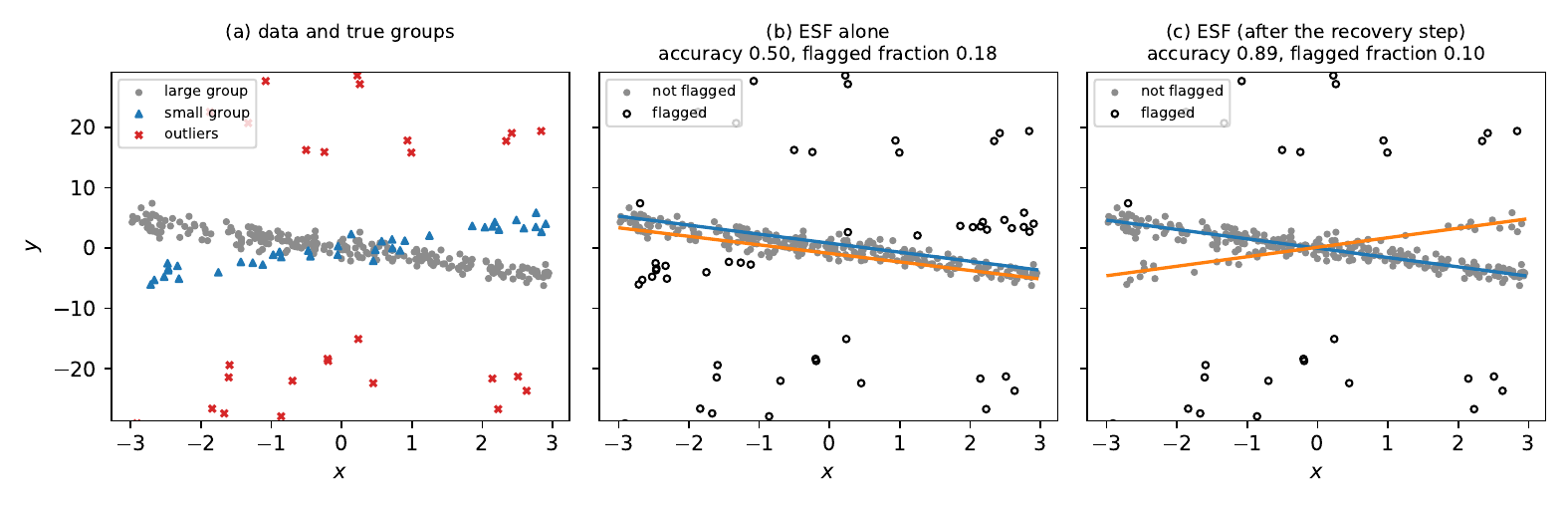}
\caption{The recovery step on one simulated data set of the third part of the study (groups of
$85\%$ and $15\%$ of the clean units, $n=300$, $\varepsilon=0.10$, $28$ outliers). (a) The true
groups. (b) The fit of ESF alone: two nearly parallel lines on the large group, with most of the
small group flagged. (c) After the recovery step. Flagged units lie more than $2.5$ robust
scales from every line.}
\label{fig:illustr}
\end{figure}

ESF flags every unit that its $K$ lines do not explain, and it cannot tell a unit that belongs
to no group from a unit of a group it has missed. The second case arises when one group is
small: the best-scoring fit may spend two nearly parallel lines on the large group and flag
most of the small group (Section~\ref{sec:further}; Figure~\ref{fig:illustr}). A trimming
method at a generous level does the same by construction. The recovery step,
Algorithm~\ref{alg:recovery}, examines the flagged set for this case. It uses only a fit
$\hat\theta$, its flagged set $\mathcal F$ and the group scales $s_k$ of the flagging rule, so
it applies after ESF and equally after TCLUST-REG with reweighting. In the
algorithm, $\mathcal F'$ is $\mathcal F$ without the units of the initial covariate screen,
which cannot form a line in the response. The weight $w$ is that of the Gaussian component in
a two-component fit to the residuals of the flagged units near the candidate line; this fit is
the \emph{peak test}.

\begin{algorithm}[tbp]
\caption{The group-recovery step, applied to a fit and its flagged set}
\label{alg:recovery}
\begin{algorithmic}[1]
\Require fit $\hat\theta$ with flags $\mathcal F$ and group scales $s_1,\dots,s_K$; noise range
$R=1.1\,(\max_iy_i-\min_iy_i)$; margin $\tfrac{d+1}{2}\log n$; restriction factor $12$ on the variances
\State stop if $|\mathcal F|>n/3$; otherwise let $\mathcal F'$ be $\mathcal F$ without the units
of the initial covariate screen, and $s=\operatorname{med}_k s_k$
\State fit a line $\beta$ to $\mathcal F'$: the line through $d$ units of $\mathcal F'$, among
$500$ random choices, that has most units of $\mathcal F'$ within $cs$, refitted three times by
least squares on the units within $cs$
\State stop unless at least $\max\{2(d+1),0.03n\}$ units of $\mathcal F'$ lie within $cs$ of
$\beta$ and they form a peak: in the model $w\,\mathcal N(0,s^{2})+(1-w)\,\mathrm{U}(-3cs,3cs)$
for the residuals from $\beta$ of the units of $\mathcal F'$ within $3cs$, the fitted $w$ is at
least $\tfrac12$
\For{$j=1,\dots,K$}
  \State $\theta^{(j)}\leftarrow$ the lines of $\hat\theta$ without line $j$, together with
  $\beta$, refitted by the reweighting step
\EndFor
\State score every fit by the log-likelihood $\ell$ of $K$ Gaussian lines plus a uniform noise
component of density $\pi_0/R$, with the scales, proportions and noise weight
$\pi_0$ read off the fit (the units of the initial covariate screen count as noise)
\State among the $\theta^{(j)}$ whose scales are all at most $\sqrt{12}\,\min_ks_k$, with $s_k$ the scales of $\hat\theta$, take the one
with the largest $\ell$; it replaces $\hat\theta$ if $\ell(\theta^{(j)})-\ell(\hat\theta)$
exceeds the margin
\State apply the step again while the fit changes, at most three times in all
\end{algorithmic}
\end{algorithm}

The steps have simple reasons. A missing group lies close to one line, so a line is sought
among the flagged units, as in RANSAC \citep{fischler1981random}. Gross outliers also contain
lines that pass near many of them, so most flagged units near a candidate must sit in a peak
of the width of the fitted groups, not in a band.
The comparison uses the likelihood of the model that the flagging rule assumes, Gaussian
groups and flagged units spread over the range of the response: two lines through one group
describe it hardly better than one line, whereas a group of $n_1$ units described as noise
instead of by its own line loses about $n_1\{\log(\pi_1R/\pi_0)-\log\sigma_1-1.42\}$ in
log-likelihood (Proposition~\ref{prop:recover-value}). The noise range $R$ is set by the
extreme responses, so gross outliers raise the value of a group; a range taken over the
unflagged units alone lost small groups (Supplement, Section~S6). The margin is a heuristic
modelled on the Bayesian information criterion. The restriction on the scales of a candidate fit is that of
TCLUST-REG: the factor $12$ bounds the ratio of the group variances, that is $\sqrt{12}$ on
the scales, as in the TCLUST-REG runs of Section~\ref{sec:simulation}, and it prevents a
candidate from absorbing the contamination into a wide group. The peak test does not exclude every band: a band of uniform noise of half-width
below about $3.4$ scales passes it (Supplement, Section~S2.3), and after a generous trimming
level the step accepted such bands at high contamination (Section~\ref{sec:gate2}). The step is
not attempted when a third of the units or more are flagged, because the fit it would examine
is then itself in doubt. The constants are fixed, not fitted to the data
(Section~\ref{sec:constants}).

\paragraph{Relation to trimmed likelihood curves.} The question the step asks, whether the
discarded units hide a group, is the one that classification trimmed likelihood curves
\citep{garciaescudero2011exploring} and the constrained-likelihood approach of
\citet{cerioli2018finding} answer for model-based clustering by fitting the model over a grid
of pairs $(K,\alpha)$ and reading the number of groups and the level off the curves. For
clusterwise regression FSDA's \texttt{tclustregIC} \citep{riani2012fsda,torti2021semiautomatic}
chooses the number of groups and the restriction factor by an information criterion at a
given level, and the level itself is monitored over a grid \citep{torti2019assessing}; the
level is thus still the analyst's choice. The recovery step answers that question locally,
post hoc and for a single candidate: it needs no grid over $\alpha$, and it works after a method that has no $\alpha$, such as ESF
or a reweighted fit; but it tests one candidate at a time, at the $K$ of the fit, and cannot
tell that the fit has one group too many or that the level was too low. When the number of
groups is in doubt, a scan over $K$ such as that of \texttt{tclustregIC} is the better tool,
and the step is then a check on what the chosen level trimmed.

\subsection{Tuning constants and defaults}
\label{sec:constants}

Supplement Table~S44 lists the constants of both algorithms, the values used in every
experiment, the range examined and the largest change in a cell's mean accuracy over that
range. Up to $20\%$ of outliers, moving one constant over its range changed a cell mean by at
most $0.05$, with four exceptions: the loss cap ($0.13$ lost at $c_{\mathrm b}=5$ in D5, while ESF
alone gains $0.24$ with a group of $15\%$ and no outliers, a gain we obtain instead through the
recovery step), the covariate quantile with high-leverage points ($0.21$), the limit on the
flagged fraction ($0.20$ at $0.25n$) and $m=12$ in every design ($0.07$ lost with four groups); $B=500$ replicates moved cell means by $-0.02$ to $+0.03$ up to $20\%$ and
by up to $-0.07$ at $30\%$.

\paragraph{The two constants that stand in for the level.} The subsample size must allow every
group $d+1$ units, so $m\ge(d+1)/\pi_{\min}$; larger values make clean replicates more accurate
but rarer, since a subsample is clean with probability close to $(1-\varepsilon)^{m}$. We used
$m=8$ for two groups and one covariate, $m=10$ when one group has $30\%$ of the units, and
$m=12$ for three groups or three covariates. The default rule is to take $m$ near
$(d+1)/\pi_{\min}$, rounded to an even number, for the lower bound $\pi_{\min}$ on the smallest
group proportion that the analyst is prepared to assume, and $m=12$ when no bound is
available. A lower bound on $\pi_{\min}$ is prior knowledge of the same kind as an upper bound
on $\varepsilon$, and the competitors are not given it, so ESF with $m=12$ in every design was
run on all data sets and appears as a row of Tables~\ref{tab:mainlow}, \ref{tab:level}
and~\ref{tab:mainhigh}. Over the $89$ cells up to $20\%$ of outliers its mean accuracy is
$0.912$ against $0.914$, with the same two cells below $0.85$, and beyond $25\%$ it is $0.813$
against $0.820$. The loss is concentrated in the four-group design ($0.88$ against $0.95$), where
$m=12$ leaves three units per group, and in clustered outliers at $\varepsilon=0.20$ ($0.75$
against $0.78$). With small groups the recovery step makes up for the smaller subsamples, which cost ESF alone
more (mean $0.864$ against $0.874$ over the $89$ cells up to $20\%$; Supplement, Tables~S12, S15, S28 and~S39). The cap of $n/3$ on the
flagged set is a guard against applying the recovery step to a fit that has broken down, not
a trimming level: moving it to $0.25n$ lost $0.16$ and $0.20$ in the two small-group designs
at $\varepsilon=0.20$, where ESF alone flags about a quarter of the units, and moving it to
$0.40n$ gained up to $0.05$ in those designs at $\varepsilon=0.30$ and moved no other cell by
more than $0.01$ (Supplement, Table~S23 and Section~S5). A higher cap is reasonable only when the
contamination is known to be below a third and a small group is suspected.

\paragraph{Other constants.} The vote weight is $\lambda=3$, with $B=100$ replicates in
$L=10$ stages (Section~\ref{sec:stage}). The cut-off $c=2.5$ is the usual value for
reweighting in robust regression \citep{rousseeuw1987robust}: at the true parameter a clean
unit is then flagged with probability $2\Phi(-2.5)\approx0.012$ (Proposition~\ref{thm:step}).
The cap $c_{\mathrm b}=3$ lies slightly above $c$, so that a unit at the flagging boundary
still contributes its full residual.

\paragraph{A default for real data.} The study used $B=100$ replicates and one seed per
data set, and on the fishery data (Section~\ref{sec:fishery}) the answer depended on the seed.
The default for real data is $B=100$ with the best of five seeds by the likelihood of the model
of the recovery step, a rule that returned the modal fit every time on the fishery data and, on
the simulated data, raised cell means up to $20\%$ of outliers and lowered one cell beyond
(Supplement, Section~S6.3). When a group is fitted exactly, as the flat tariff of
Section~\ref{sec:taxi}, the likelihood rewards any line through tied points and the seeds
should be compared by eye. The flagged fraction $\hat\alpha$ is the check on the result. A
value well below any plausible contamination means that the outliers have been absorbed into a
fitted line and the fit has broken down (Section~\ref{sec:high}). A value well above it,
as on the tone and taxi data, means that the errors are not close to Gaussian or that a group's
covariates are compact; $\hat\alpha$ is then not an estimate of the contamination.

\section{Theoretical results}
\label{sec:theory}

Three partial results follow; their precise statements, proofs and scope are in Section~S2 of
the Supplement, and none covers the passage from one stage of ESF to the next, the weighted
vote, or the search for the candidate line. Throughout, $O\subseteq\{1,\dots,n\}$ is a fixed set of
contaminated units whose values are arbitrary, and a subsample is \emph{clean} if it contains
no unit of $O$.

\paragraph{One stage of the sequential draw.}
\label{sec:stage}
Let $E_\ell$ be the set of units not flagged after stage $\ell-1$, from which stage $\ell$
draws, and $\varepsilon_\ell=|E_\ell\cap O|/|E_\ell|$.

\begin{proposition}[informal statement; the precise version is in Supplement~S2.0]
\label{prop:clean}
Conditionally on the data and on everything drawn before stage $\ell$, the subsamples of stage
$\ell$ are independent and uniform on the admissible subsamples of size $m$ of $E_\ell$, each
replicate is clean with probability $\pi_\ell=p_\ell a_\ell$, where
$p_\ell\le(1-\varepsilon_\ell)^{m}$ is the probability that a uniform subsample of $E_\ell$ is
clean and $a_\ell$ corrects for admissibility, and the stage contains a clean replicate
with probability $1-(1-\pi_\ell)^{B/L}$.
\end{proposition}

The result is exact and covers one stage: a clean replicate becomes likelier as the flags
remove outliers from $E_\ell$, but nothing here says that earlier stages do so. With $m=8$ the
first stage contains a clean replicate with probability about $0.84$ at $\varepsilon=0.20$ and
$0.45$ at $0.30$, and with $m=12$ about $0.51$ and $0.13$, which is where ESF breaks down.

\paragraph{The final reweighting step.}
\label{sec:reweight}
The reweighting step takes a parameter $\theta$, recomputes the flags and refits each group on
its unflagged units. Let the units be independent with law $P=(1-\varepsilon)P_0+\varepsilon C$,
where under $P_0$ a unit of group $k$, drawn with probability $\pi_k>0$, has
$y=x^{\top}\theta^{0}_k+\sigma_ke$ with $e\sim\mathcal N(0,1)$ independent of $x$, and every
unit of $C$ lies beyond $c'>c$ scales from the nearest line; let $\eta$ be the probability that
a clean unit is nearer to another true line than to its own.

\begin{proposition}[informal statement; the precise version is Theorem~S2 of the Supplement]
\label{thm:step}
Apply the map at $\theta=\theta^{0}$. Under mild regularity conditions, almost surely as
$n\to\infty$, the scale of group $k$ converges to a limit $s_k$ with $s_k/\sigma_k=1+O(\nu_k)$,
$\nu_k$ being of the order of $\eta+\varepsilon$ divided by the mass assigned to line $k$; the
flagged fraction converges to $\varepsilon+(1-\varepsilon)\kappa$, where $\kappa$, the fraction
of clean units flagged, lies within $\eta$ of $\sum_k\pi_k2\Phi(-cs_k/\sigma_k)$; and the
refitted coefficients converge to a limit that depends on the contamination only through the
$s_k$ and lies within $O(\eta)$ of $\theta^{0}_k$ when the covariates are bounded. Without
overlap or contamination, $\theta^{0}$ is a fixed point of the map and the flagged fraction
tends to $2\Phi(-c)$.
\end{proposition}

This is a law of large numbers for one application of the map at a fixed input: it says what
the flagged fraction estimates under Gaussian errors, $\varepsilon$ plus a fraction of clean
units close to $2\Phi(-c)\approx0.012$ when the groups overlap little, and that contamination
beyond the cut-off is removed entirely. It does not cover four things: a small group under
moderate contamination, where the condition $\nu_k<\tfrac12$ of Theorem~S2 fails; the input
that ESF delivers to the map; the repeated application of the map in the code; and the flagged
fraction reported at the refitted parameter (Supplement, Section~S2.2).

\paragraph{The group-recovery step.}
\label{sec:recovery-theory}
The recovery step scores a \emph{fit} $F$, a set of lines $\theta_k$ with scales $s_k$ and
proportions $\pi_k$ and a noise weight $\pi_0>0$ summing to one, by
$\ell_n(F)=\sum_{i\le n}\log f_F(u_i)$, where
$f_F(u)=\sum_{k}(\pi_k/s_k)\,\varphi\{(y-x^{\top}\theta_k)/s_k\}+\pi_0/R$. For given
parameters, $\ell_n(F)/n$ converges almost surely to $L(F)=\E_P\log f_F(u)$; the proposition
values, in $L$, the two operations by which a candidate differs from the original fit.

\begin{proposition}[informal statement; the precise version is in Supplement~S2.3]
\label{prop:recover-value}
(i) Let $G$ be an event of probability $\pi_\ast$ on which $y=x^{\top}\theta_\ast+\sigma_\ast e$
with $e\sim\mathcal N(0,1)$ independent of $x$, let $F$ have noise weight $\pi_0>\pi_\ast$, and
let $F'$ be $F$ with the line $(\theta_\ast,\sigma_\ast,\pi_\ast)$ added and the noise weight
lowered to $\pi_0-\pi_\ast$. Then
\begin{equation}
L(F')-L(F)\;\ge\;\pi_\ast\Bigl\{\log\frac{\pi_\ast R}{\pi_0}-\log\sigma_\ast
-\tfrac12\log(2\pi e)-\xi\Bigr\}-\nu\log\frac{\pi_0}{\pi_0-\pi_\ast},
\label{eq:recover-gain}
\end{equation}
where $\xi\ge0$ is small when $G$ lies a few scales from every line of $F$, and
$\nu\le1-\pi_\ast$ is the noise mass that $F$ attributes to the units outside $G$. (ii) If $F$
contains two lines through one group of probability $\pi_S$ and scale $\sigma_S$, at intercepts
$\delta$ apart, and $F''$ replaces them by the line through the group, then
$|L(F'')-L(F)|\le(\varphi(1)a^{2}+\tfrac14a^{4})\pi_S$ with $a=\delta/(2\sigma_S)$.
\end{proposition}

The leading term of \eqref{eq:recover-gain} is the value of a group: per unit, the log of
the ratio of its peak density $\pi_\ast/\sigma_\ast$ to the noise density $\pi_0/R$, less the
entropy $\tfrac12\log(2\pi e)\approx1.42$ of a unit normal error. With $R/\sigma_\ast=40$,
$\pi_\ast=0.1$, $\pi_0=0.2$ and $\xi\approx0$, the value is $1.58$ nats per unit of the group,
that is $0.158n$ in all. With these values the loss term is $\nu\log2$: for a fit whose noise
mass off $G$ is about $0.1$, the weight $\pi_0=0.2$ less the group's $0.1$, it is about $0.069n$
and the gain at least $0.09n$ against a margin $\tfrac{d+1}{2}\log n$ of $8.6$ for $d=2$,
$n=300$. The bound $\nu\le1-\pi_\ast$ alone gives a loss of up to $0.62n$, which exceeds the
value, so the inequality is informative only for such a fit. Part (ii), an idealisation with two parallel lines at intercepts $\delta$ apart, says that
removing one of two lines through one group is nearly free, so a candidate that arises by the two operations is accepted
for all large $n$ when the right side of \eqref{eq:recover-gain} exceeds $\varphi(1)a^{2}\pi_S$
and its scales are admissible. The result is for fixed parameters, which the step reads off
the data, and it does not cover the reweighting refit of the candidate. The peak test, analysed
in Supplement~S2.3, fails a band of outliers that fills its window and passes one of half-width
below about $3.4$ scales.

\section{Simulation study}
\label{sec:simulation}

\subsection{Design}
\label{sec:design}

\paragraph{Data.} Each data set has $n=300$ units, a covariate uniform on $[-3,3]$ and standard
normal errors. The eight designs are: D1, two crossing lines with equal weights and noise;
D2, three parallel lines three noise scales apart ($m=12$); D3, as D1 with weights $0.7$ and
$0.3$ ($m=10$); D4, as D1 with noise scales $0.5$ and $1.5$; D5, as D1 with high-leverage
outliers; D6, as D1 with uniform outliers; D7, as D1 with two further covariates of slope $1$
($m=12$); D8, two parallel lines six noise scales apart; $m=8$ unless stated. In D1, D3, D4
and D6 the two lines cross at the origin with slopes $\pm1.5$, so the groups overlap near the
crossing and even the Bayes classifier with known parameters labels about $9\%$ of the units
wrongly; an accuracy of about $0.91$ is the ceiling there, and essentially one in D8. Each
unit is replaced by an outlier independently with probability $\varepsilon$, so the realised
fraction of outliers varies around $\varepsilon$. Outliers in the response are shifted up or
down by a uniform amount between $15$ and $30$ times the average noise scale; high-leverage
outliers (D5) have their covariate redrawn uniformly on $[6,9]$, outside the design range, and
their response uniformly on $[-3,3]$; uniform outliers (D6) are spread over the bounding box
of the clean data enlarged by $20\%$, away from the true lines.

\paragraph{Competitors.} The main rival is TCLUST-REG \citep{garciaescudero2010robust}, from the
FSDA toolbox \citep{riani2012fsda}, run unmodified under GNU Octave (a check in MATLAB on $80$ data sets is in Supplement, Section~S1) with restriction factor $12$,
$1000$ random starts, $50$ refinement steps and no second-level trimming in the covariates, at
the two conservative levels $\alpha=0.25$ and $0.40$, followed by reweighting and then by the
recovery step. The reweighting attached to TCLUST-REG is our own rule, the map of
Section~\ref{sec:reweight} applied twice with $c=2.5$ and a median absolute deviation per group,
which carries the idea of \citet{dotto2018reweighting} to regression; it is not the incremental
scheme of that paper, which is examined separately (Supplement, Section~S6). ``TCLUST-REG,
reweighted'' denotes this construction throughout. The Supplement (Tables~S4 to~S11, S37
and~S38) holds the rows of a shorter search ($300$ starts, $10$ refinement steps,
$\alpha\in\{0.05,0.10,0.15,0.25,0.40\}$), which understated TCLUST-REG with three groups,
and those of the trimmed cluster-weighted model \citep{garciaescudero2017robust} and the
trimmed likelihood estimator \citep{neykov2007robust} at the same five levels. CTLE
\citep{chang2020componentwise} and the bisquare mixture of regressions \citep{bai2012robust}
were the competitors without a level from published software; they come from the R package
RobMixReg \citep{cao2026robmixreg}, as do the two trimmed methods. Four further competitors without a level were added after the first part had
been analysed: the mixture with Laplace errors of \citet{song2014robust}; the mixture of
contaminated normal regressions \citep{punzo2017robust,mazza2020mixtures}; a mixture with a
uniform noise component in the manner of \citet{banfield1993model}, the last two our
implementations; and sequential RANSAC \citep{fischler1981random}, which extracts the $K$
lines one at a time and is given the true noise scale, so it is an oracle benchmark, not a
competitor that can be run on data. Least squares with $50$ random starts is a reference
without protection (software and settings: Supplement, Section~S1). The monitoring approach
of \citet{torti2019assessing} leaves the choice of the level to the analyst; it enters the
comparison only through two automatic rules for reading a level off the monitoring path
(Section~\ref{sec:measures}).

\paragraph{Criteria.} Every method is given the true number of groups $K$ and returns $K$
coefficient vectors, from which every clean unit receives the label of its nearest fitted
line. The \emph{accuracy} is the fraction of clean units whose label agrees with the true one,
maximised over relabellings; trimmed and flagged units are treated alike, and a failed fit counts
as accuracy zero (failures were rare except for CTLE; Supplement, Section~S5). This labelling
ignores the estimated scales and weights, so it is not the Bayes rule of any method where the
groups differ in scale, as in D4, and it does not penalise trimming or flagging clean units as
such. The coefficient error, the flagged fraction and the computing time are therefore
reported as well (Section~\ref{sec:measures}), and so are the adjusted Rand index on all
units, with the outliers as a class of their own, and a misclassification rate that counts a
flagged clean unit as an error. By these two criteria the main pipelines rank as they do by accuracy: up to
$\varepsilon=0.20$ the mean adjusted Rand index is $0.719$ for ESF and $0.725$ and $0.715$ for
TCLUST-REG at levels $0.25$ and $0.40$ with reweighting and the recovery step, beyond it $0.660$
against $0.519$ and $0.714$ (Supplement, Table~S45). Labelling the clean units by TCLUST-REG's
own assignment instead of the nearest line changed its mean accuracy in D2, D4 and with four
groups by $-0.03$ to $+0.02$ (Supplement, Table~S49), so the criterion does not drive the
comparison.

\paragraph{Analysis plans.} The study was run in parts, and before each part a written plan
fixed the comparison to be made and its thresholds (Plans~1 to~5; Supplement, Table~S47, with
the digests of the plan files in Table~S34); the third part is descriptive. Four things
reported below were decided after all the plans: the main rival pipeline, TCLUST-REG with the
long search, reweighting and the recovery step; the three mixtures and sequential RANSAC; the
best-of-five rule of Section~\ref{sec:constants}; and a guard in the scoring scale of ESF for
groups fitted exactly (Section~\ref{sec:taxi}), which changed nothing on the $40$ simulated
data sets on which we checked it.

\subsection{Up to 20\% of outliers}
\label{sec:low}

Table~\ref{tab:mainlow} shows, for each design and method, the lowest mean accuracy over
$\varepsilon\in\{0,0.05,0.10,0.20\}$; if a method works only at the right trimming level, or
only with equal groups, at least one entry in its row is low. Supplement Table~S40 gives the
TCLUST-REG rows cell by cell, with the paired differences from ESF.

\begin{table}[!tb]
\centering
\footnotesize
\caption{Worst mean accuracy on clean units (the lowest of the cell means) with up to $20\%$ of outliers ($\varepsilon\in\{0,0.05,0.10,0.20\}$), by design. D1 to D8: first part ($50$ data sets per cell). $K=4$, small groups (80/20 and 85/15, including the fresh data of Plan~4), clustered outliers and other designs (sample size, moderate outliers, dependent covariates, $t_3$, skewed and heteroscedastic errors): third part and Plans~4 and~5. Column entries are minima over different sets of $\varepsilon$: D1 to D8 over the four values above, the other columns over the values of $\varepsilon\le0.20$ run in the third part and Plans~4 and~5. TCLUST-REG was run with $1000$ random starts and $50$ refinement steps (the rows with the level $0.30$ and with equal weights were run after the others); ``reweighted'' is our two-pass rule ($c=2.5$, median absolute deviation per group). Sequential RANSAC is given the true noise scale and is an oracle benchmark, not a competitor that can be run on data. CTLE and the mixtures were run with the starts of Supplement Section~S1. In D5, TCLUST-REG at level $0.25$ with $5\%$ second-level trimming in the covariates and reweighting ($300$ starts) reaches $0.86$ (Supplement, Table~S37). Values below $0.85$ in italics.}
\label{tab:mainlow}
\resizebox{\textwidth}{!}{%
\begin{tabular}{lcccccccccccc}
\toprule
Method & D1 & D2 & D3 & D4 & D6 & D7 & D8 & D5 & $K=4$ & Small & Clust. & Other \\
\midrule
ESF (with the recovery step) & 0.91 & 0.89 & 0.90 & 0.90 & 0.89 & 0.91 & 1.00 & 0.85 & 0.95 & \textit{0.83} & \textit{0.78} & 0.90 \\
ESF with $m=12$ in every design & 0.91 & 0.90 & 0.89 & 0.90 & 0.89 & 0.91 & 1.00 & 0.86 & 0.88 & \textit{0.83} & \textit{0.75} & 0.90 \\
ESF alone & 0.91 & 0.89 & 0.89 & 0.90 & 0.90 & 0.91 & 1.00 & 0.85 & 0.95 & \textit{0.60} & \textit{0.70} & 0.88 \\
\addlinespace
TCLUST-REG, $\alpha=0.25$, reweighted, then the recovery step & 0.91 & 0.89 & 0.91 & 0.90 & 0.91 & 0.91 & 1.00 & \textit{0.83} & 0.91 & 0.90 & 0.90 & 0.91 \\
TCLUST-REG, $\alpha=0.40$, reweighted, then the recovery step & 0.91 & \textit{0.83} & 0.91 & 0.90 & 0.91 & 0.91 & 1.00 & 0.87 & \textit{0.81} & 0.85 & 0.91 & 0.89 \\
TCLUST-REG, $\alpha=0.30$, reweighted, then the recovery step & 0.91 & 0.87 & 0.91 & 0.90 & 0.91 & 0.91 & 1.00 & 0.87 & 0.88 & 0.89 & 0.91 & 0.90 \\
TCLUST-REG, $\alpha=0.25$, equal weights, reweighted, then the recovery step & 0.91 & 0.90 & 0.91 & 0.90 & 0.91 & 0.91 & 1.00 & 0.86 & 0.96 & 0.89 & 0.90 & 0.90 \\
TCLUST-REG, $\alpha=0.25$, reweighted & 0.91 & 0.89 & 0.89 & 0.90 & 0.91 & 0.91 & 1.00 & \textit{0.83} & 0.91 & \textit{0.60} & 0.90 & 0.88 \\
TCLUST-REG, $\alpha=0.40$, reweighted & 0.91 & \textit{0.83} & \textit{0.66} & 0.90 & 0.91 & 0.91 & 1.00 & 0.87 & \textit{0.81} & \textit{0.56} & 0.91 & \textit{0.60} \\
\addlinespace
CTLE & 0.91 & \textit{0.59} & 0.87 & 0.85 & 0.87 & 0.89 & \textit{0.72} & \textit{0.73} & \textit{0.55} & \textit{0.64} & 0.86 & \textit{0.81} \\
Bisquare mixture & 0.89 & \textit{0.56} & 0.88 & 0.90 & 0.90 & 0.90 & \textit{0.80} & \textit{0.55} & \textit{0.47} & \textit{0.81} & \textit{0.57} & 0.85 \\
Laplace mixture & 0.91 & \textit{0.61} & 0.91 & 0.91 & 0.91 & 0.91 & 0.96 & \textit{0.58} & \textit{0.60} & 0.88 & \textit{0.63} & 0.90 \\
Contaminated normal mixture & 0.91 & \textit{0.52} & 0.91 & 0.91 & 0.91 & 0.91 & 1.00 & \textit{0.83} & \textit{0.65} & 0.91 & \textit{0.58} & 0.89 \\
Noise-component mixture & 0.91 & \textit{0.75} & 0.91 & 0.90 & 0.91 & 0.90 & \textit{0.81} & \textit{0.53} & \textit{0.65} & \textit{0.64} & \textit{0.61} & 0.88 \\
Sequential RANSAC (oracle: true noise scale given) & 0.91 & \textit{0.69} & 0.91 & 0.90 & 0.91 & 0.91 & 0.97 & \textit{0.54} & \textit{0.71} & 0.91 & \textit{0.84} & 0.90 \\
Least squares (50 starts) & \textit{0.53} & \textit{0.37} & \textit{0.70} & \textit{0.52} & \textit{0.58} & \textit{0.53} & \textit{0.53} & \textit{0.53} & \textit{0.49} & \textit{0.80} & \textit{0.61} & \textit{0.51} \\
\bottomrule
\end{tabular}}
\end{table}

The rows of TCLUST-REG show what a generous level costs and what the recovery step returns. At
level $0.40$ with reweighting, TCLUST-REG trims away the smaller group whenever the groups are
unequal: its worst accuracy is $0.66$ in D3, where the smaller group holds $30\%$ of the units,
and $0.56$ with groups of $15\%$ and $20\%$. The recovery step returns these groups ($0.91$ and
$0.85$) but does not repair three and four groups, where the level discards part of several
groups at once ($0.83$ and $0.81$ with or without the step). At level $0.25$ with reweighting
and the recovery step, TCLUST-REG is at $0.89$ or more in every design except D5 ($0.83$);
without the step it loses the small groups at $\varepsilon=0$ ($0.60$). TCLUST-REG was run
without second-level trimming in the covariates, so in D5 the high-leverage points can be
trimmed only through their residuals, hence the $0.83$; the adaptive form of that trimming did
not help either (Supplement, Section~S6.1).
ESF is at $0.85$ or more in every design except clustered outliers at $\varepsilon=0.20$
($0.78$) and groups of $80\%$ and $20\%$ with three covariates at $\varepsilon=0.20$ ($0.83$).
With four groups it reaches $0.95$, against $0.91$ for the level $0.25$ and $0.81$ for the level
$0.40$ with estimated weights, and is matched only by TCLUST-REG at level $0.25$ with equal
weights ($0.96$). CTLE and the four mixtures all fall below $0.85$ with three or four groups and with
high-leverage points. With three groups CTLE and the Laplace and bisquare mixtures are below least
squares with $50$ starts even without outliers ($0.78$, $0.66$ and $0.72$ against $0.90$), so
their default search is part of the failure; a start from least squares raised the two mixtures
we implemented to $0.89$ with three groups and $0.95$ with four without outliers, but not with
outliers, and $100$ starts did not help (Supplement, Tables~S21 and~S27). Sequential RANSAC, although given the true noise scale, falls to
$0.69$ with three parallel groups and $0.71$ with four.

\begin{table}[htbp]
\centering
\small
\caption{The choice of a trimming level: mean over cells of the mean accuracy on clean units, the
number of cells with a mean below $0.85$, and the worst cell, for all $123$ cells of the study
(Plans~1 to~5 and the third part; the $\varepsilon=0.30$ cells of the first part are excluded,
as the second part replaces them). ``Better'' and ``worse'' of the two levels are chosen in each
cell after the run, between TCLUST-REG at $0.25$ and at $0.40$, each with reweighting and the
recovery step. The last four rows were run after the others: the levels $0.30$ and $0.35$, and
the levels $0.25$ and $0.40$ with equal weights (option \texttt{equalweights} of FSDA), all with
the long search.}
\label{tab:level}
\resizebox{\textwidth}{!}{\begin{tabular}{lcccccc}
\toprule
 & \multicolumn{3}{c}{$\varepsilon\le0.20$ (89 cells)} & \multicolumn{3}{c}{$\varepsilon\ge0.25$ (34 cells)} \\
\cmidrule(lr){2-4}\cmidrule(lr){5-7}
Method & Mean & Below $0.85$ & Worst & Mean & Below $0.85$ & Worst \\
\midrule
ESF & 0.914 & 2 & 0.78 & 0.820 & 13 & 0.53 \\
ESF, $m=12$ in every design & 0.912 & 2 & 0.75 & 0.813 & 13 & 0.53 \\
TCLUST-REG, $\alpha=0.25$, rw, recovery & 0.917 & 1 & 0.83 & 0.693 & 26 & 0.51 \\
TCLUST-REG, $\alpha=0.40$, rw, recovery & 0.909 & 4 & 0.81 & 0.863 & 8 & 0.53 \\
Better of the two levels, chosen afterwards & 0.918 & 0 & 0.87 & 0.865 & 7 & 0.53 \\
Worse of the two levels & 0.908 & 5 & 0.81 & 0.691 & 27 & 0.51 \\
TCLUST-REG, $\alpha=0.25$, rw, no recovery & 0.890 & 11 & 0.60 & 0.695 & 26 & 0.51 \\
TCLUST-REG, $\alpha=0.40$, rw, no recovery & 0.827 & 31 & 0.56 & 0.869 & 7 & 0.54 \\
\addlinespace
TCLUST-REG, $\alpha=0.30$, rw, recovery & 0.915 & 0 & 0.87 & 0.794 & 17 & 0.53 \\
TCLUST-REG, $\alpha=0.35$, rw, recovery & 0.912 & 4 & 0.83 & 0.851 & 10 & 0.53 \\
TCLUST-REG, $\alpha=0.25$, equal weights, rw, recovery & 0.919 & 0 & 0.86 & 0.769 & 24 & 0.53 \\
TCLUST-REG, $\alpha=0.40$, equal weights, rw, recovery & 0.907 & 4 & 0.79 & 0.856 & 10 & 0.54 \\
\bottomrule
\end{tabular}}
\end{table}

Table~\ref{tab:level} sets the three pipelines against the choice an analyst has to make.
With up to $20\%$ of outliers, the better of the two levels, chosen after the run, never falls
below $0.85$, while the worse one does so in five cells. ESF, which chooses nothing, is within
$0.03$ of the better of the levels $0.25$ and $0.40$ in $83$ of the $89$ cells, above it by
more than $0.03$ in two (four groups at $\varepsilon=0.10$ and $0.20$) and below it by more
than $0.03$ in four: clustered
outliers and the three small-group designs of Plans~3 and~4 at $\varepsilon=0.20$, with gaps of
$0.03$ to $0.125$. Without the recovery step, the level $0.40$ falls below $0.85$ in $31$
cells, $27$ of them with unequal or small groups. These counts describe design points chosen
to probe weaknesses, not frequencies in any population of data sets. With $200$ data sets in
each of the eleven cells that decide the comparison ($150$ added; Supplement, Table~S50), the
standard errors of the paired differences fall to about $0.01$ and the conclusions stand: with
four groups ESF is ahead of TCLUST-REG at levels $0.25$ and $0.30$ by $0.03$ to $0.06$ and on a
par with the level $0.25$ with equal weights, and with a small group or clustered outliers at
$\varepsilon=0.20$ it is behind the level $0.25$ by $0.03$ to $0.09$.

The levels in between and equal weights trade one range of contamination for the other
(Table~\ref{tab:level}, last four rows). With reweighting and the recovery step, the level
$0.30$ falls below $0.85$ in no cell up to $20\%$ of outliers and in $17$ of the $34$ beyond,
the level $0.35$ in four and ten. Equal weights make the level $0.25$ the most accurate pipeline
up to $20\%$ (mean $0.919$, no cell below $0.85$; ESF is within $0.03$ of it in $85$ of the
$89$ cells and below it by more in four) but leave $24$ cells below $0.85$ beyond. They
change the summaries of the level $0.40$ little, although cell by cell they gain $0.01$ to
$0.05$ with three groups and $0.04$ to $0.05$ with four and lose up to $0.20$ with unequal or small groups. The two
pipelines that do best up to $20\%$ appear as rows of Table~\ref{tab:mainlow}: the level
$0.30$ has no entry below $0.85$ and reaches $0.88$ with four groups, where the level $0.25$
with equal weights reaches $0.96$ and ESF $0.95$. Over all $123$ cells the level $0.40$ with reweighting and the
recovery step has the highest mean ($0.897$), then the level $0.35$ ($0.895$), the level $0.40$
with equal weights ($0.893$) and ESF ($0.888$). By mean accuracy and by the number of cells within $0.03$ of ESF up to $20\%$, the level
$0.35$ is the closest to a level that serves both ranges: up to $20\%$ of outliers it is within $0.03$ of ESF in $83$ of the $89$ cells
(mean $0.912$ against $0.914$), ahead in two and behind in four, three of them with four groups,
where it loses $0.11$ to $0.13$; beyond $25\%$ it is ahead of ESF in $12$ of the $34$ cells and behind in four.
Table~\ref{tab:level} is descriptive: these levels were run after the analysis plans.

The criterion of Plan~1, in the seven designs without high-leverage points at
$\varepsilon\le0.20$, $28$ cells, was met in every cell by ESF alone, the largest shortfall
against the best competitor being $0.029$ (standard error $0.014$), and again by ESF with the
recovery step under Plan~3 (Supplement, Table~S1 and Section~S6). Its second criterion, that every
fixed-level competitor trail ESF alone by at least $0.05$ in some cell, was not met: TCLUST-REG at level
$0.25$ with reweighting was never more than $0.033$ below it, so this competitor was tested
again, on new data, at higher contamination.

\subsection{Beyond 25\% of outliers}
\label{sec:high}

Plan~2 asked whether TCLUST-REG at level $0.25$ with reweighting fails once the contamination
exceeds its level while ESF alone does not. The answer was yes, by a wide margin and by construction:
at $\varepsilon=0.30$ and $0.35$ this competitor fell to accuracies between $0.53$ and $0.62$
in D1, D4, D7 and D8, with the short search and with the long one (Table~\ref{tab:mainhigh}),
where ESF alone stayed above $0.87$.

\begin{table}[!tb]
\centering
\footnotesize
\caption{Worst mean accuracy on clean units with $25\%$ to $35\%$ of outliers. D1 to D8: second part, new data ($100$ data sets per cell); the other columns: third part at $\varepsilon=0.30$ only ($50$ data sets per cell), so column entries are minima over different sets of $\varepsilon$. Rows and conventions as in Table~\ref{tab:mainlow}. Values below $0.85$ in italics.}
\label{tab:mainhigh}
\resizebox{\textwidth}{!}{%
\begin{tabular}{lcccccccccccc}
\toprule
Method & D1 & D2 & D3 & D4 & D6 & D7 & D8 & D5 & $K=4$ & Small & Clust. & Other \\
\midrule
ESF (with the recovery step) & 0.91 & \textit{0.55} & \textit{0.84} & 0.90 & \textit{0.58} & 0.88 & 1.00 & \textit{0.53} & \textit{0.72} & \textit{0.75} & \textit{0.55} & 0.87 \\
ESF with $m=12$ in every design & 0.91 & \textit{0.56} & \textit{0.83} & 0.89 & \textit{0.55} & 0.87 & 0.98 & \textit{0.53} & \textit{0.77} & \textit{0.66} & \textit{0.56} & 0.87 \\
ESF alone & 0.91 & \textit{0.56} & \textit{0.84} & 0.90 & \textit{0.59} & 0.88 & 1.00 & \textit{0.53} & \textit{0.73} & \textit{0.73} & \textit{0.54} & 0.87 \\
\addlinespace
TCLUST-REG, $\alpha=0.25$, reweighted, then the recovery step & \textit{0.53} & \textit{0.51} & \textit{0.70} & \textit{0.53} & \textit{0.69} & \textit{0.53} & \textit{0.53} & \textit{0.53} & \textit{0.74} & \textit{0.82} & \textit{0.59} & \textit{0.52} \\
TCLUST-REG, $\alpha=0.40$, reweighted, then the recovery step & 0.91 & \textit{0.76} & 0.87 & 0.90 & \textit{0.72} & 0.90 & 1.00 & \textit{0.53} & \textit{0.76} & \textit{0.82} & \textit{0.70} & 0.91 \\
TCLUST-REG, $\alpha=0.30$, reweighted, then the recovery step & \textit{0.60} & \textit{0.61} & \textit{0.74} & \textit{0.63} & \textit{0.73} & \textit{0.59} & \textit{0.60} & \textit{0.53} & \textit{0.77} & 0.86 & \textit{0.62} & \textit{0.77} \\
TCLUST-REG, $\alpha=0.25$, equal weights, reweighted, then the recovery step & \textit{0.59} & \textit{0.66} & \textit{0.71} & \textit{0.57} & \textit{0.74} & \textit{0.57} & \textit{0.64} & \textit{0.53} & \textit{0.78} & \textit{0.82} & \textit{0.61} & \textit{0.75} \\
TCLUST-REG, $\alpha=0.25$, reweighted & \textit{0.53} & \textit{0.51} & \textit{0.70} & \textit{0.53} & \textit{0.78} & \textit{0.53} & \textit{0.53} & \textit{0.53} & \textit{0.74} & \textit{0.82} & \textit{0.55} & \textit{0.52} \\
TCLUST-REG, $\alpha=0.40$, reweighted & 0.91 & \textit{0.76} & 0.87 & 0.90 & 0.90 & 0.90 & 1.00 & \textit{0.54} & \textit{0.76} & \textit{0.81} & \textit{0.68} & 0.91 \\
\addlinespace
CTLE & \textit{0.65} & \textit{0.40} & \textit{0.65} & \textit{0.47} & \textit{0.67} & \textit{0.55} & \textit{0.66} & \textit{0.51} & \textit{0.47} & \textit{0.62} & \textit{0.67} & \textit{0.69} \\
Bisquare mixture & \textit{0.73} & \textit{0.42} & \textit{0.82} & \textit{0.76} & \textit{0.83} & \textit{0.63} & \textit{0.75} & \textit{0.53} & \textit{0.46} & \textit{0.80} & \textit{0.57} & \textit{0.68} \\
Laplace mixture & 0.91 & \textit{0.48} & 0.90 & 0.91 & 0.90 & 0.90 & 0.94 & \textit{0.53} & \textit{0.53} & 0.86 & \textit{0.61} & 0.90 \\
Contaminated normal mixture & 0.90 & \textit{0.37} & 0.91 & 0.90 & \textit{0.67} & \textit{0.65} & 0.85 & \textit{0.61} & \textit{0.48} & 0.88 & \textit{0.53} & \textit{0.71} \\
Noise-component mixture & \textit{0.55} & \textit{0.40} & \textit{0.72} & \textit{0.55} & 0.87 & \textit{0.54} & \textit{0.53} & \textit{0.52} & \textit{0.46} & \textit{0.83} & \textit{0.57} & \textit{0.56} \\
Sequential RANSAC (oracle: true noise scale given) & 0.91 & \textit{0.64} & 0.91 & 0.91 & \textit{0.73} & 0.91 & 0.98 & \textit{0.53} & \textit{0.71} & 0.89 & \textit{0.63} & 0.90 \\
Least squares (50 starts) & \textit{0.53} & \textit{0.37} & \textit{0.70} & \textit{0.53} & \textit{0.53} & \textit{0.53} & \textit{0.53} & \textit{0.53} & \textit{0.47} & \textit{0.81} & \textit{0.62} & \textit{0.51} \\
\bottomrule
\end{tabular}}
\end{table}

\begin{figure}[tbp]
\centering
\includegraphics[width=0.9\textwidth]{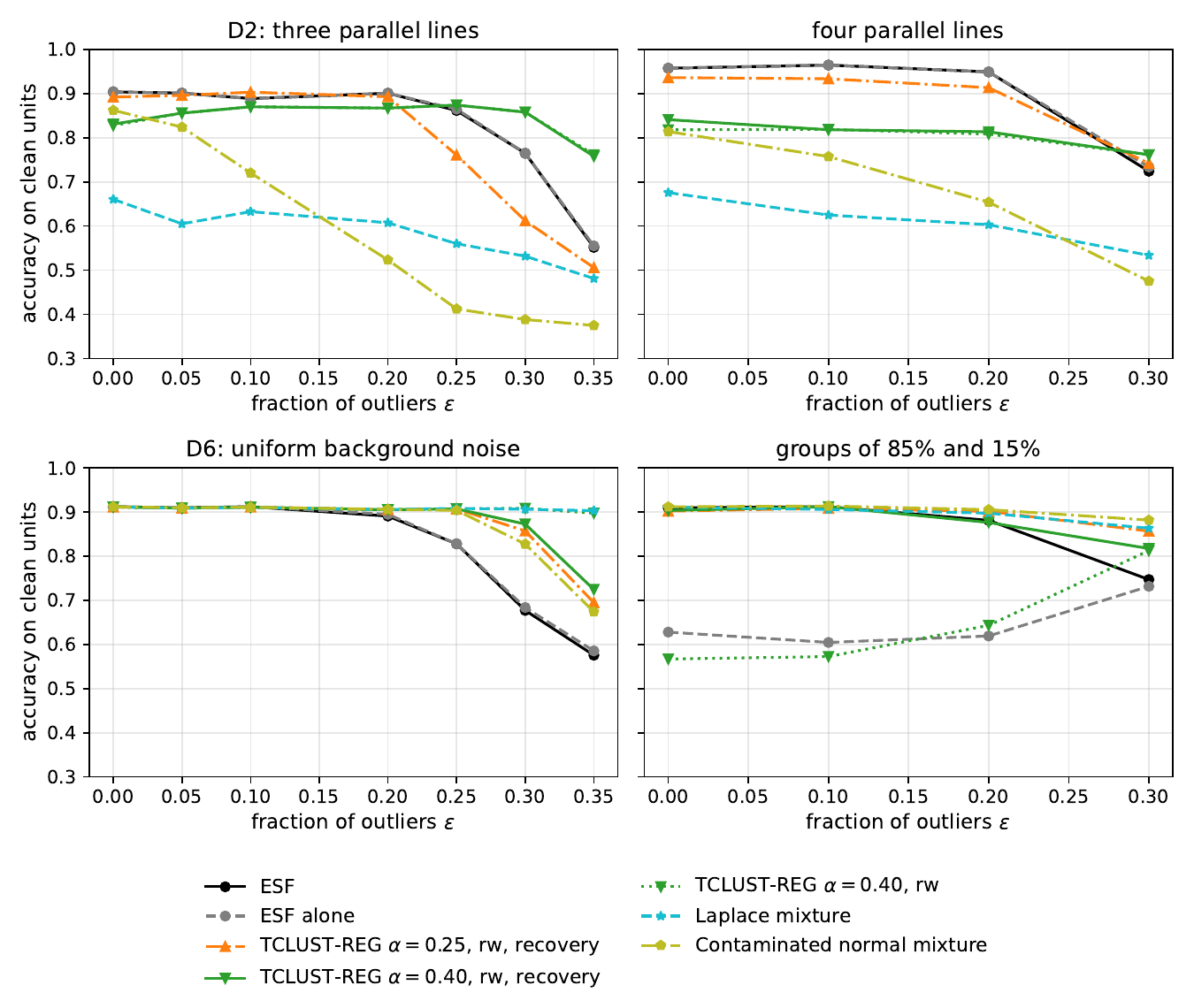}
\caption{Mean accuracy on clean units against the fraction of outliers in four designs. D2 and D6:
first part ($\varepsilon\le0.20$, $50$ data sets each) and second part ($\varepsilon\ge0.25$, $100$
data sets each, new data); four groups and groups of $85\%$ and $15\%$: third part ($50$ data sets
each), which was run at $\varepsilon\le0.30$ only. TCLUST-REG with $1000$ starts; rw: reweighted
by our rule; recovery: followed by the recovery step.}
\label{fig:curves}
\end{figure}

The level $0.40$ is the more robust choice from $25\%$ of outliers. With reweighting and the
recovery step it is ahead of ESF by more than $0.03$ in $13$ of the $34$ cells and never behind
by as much (Table~\ref{tab:level}; Figure~\ref{fig:curves}). ESF breaks down in four
situations. With uniform background noise (D6) it falls to $0.83$ at $\varepsilon=0.25$ and
$0.58$ at $0.35$, while TCLUST-REG at level $0.40$ with reweighting and the Laplace mixture stay
at $0.90$, and the recovery step, applied after TCLUST-REG, lowers this to $0.72$
(Section~\ref{sec:gate2}). With three groups (D2) it falls to $0.76$ at $\varepsilon=0.30$ and
$0.55$ at $0.35$, against $0.86$ and $0.76$ for the level $0.40$, and with four groups at
$0.30$ they reach $0.72$ and $0.76$. With high-leverage outliers (D5) it falls to $0.64$ at
$\varepsilon=0.25$ and about $0.53$ beyond, where no method is reliable, the best of
Supplement Table~S11 being the trimmed cluster-weighted model at level $0.40$ ($0.83$ at
$\varepsilon=0.25$, $0.70$ at $0.35$). With a small group or clustered outliers at $30\%$ it reaches $0.75$ and $0.55$,
against $0.82$ to $0.87$ and $0.70$ for the level $0.40$. In D1, D4, D7 and D8 it holds up to
$35\%$ of outliers. The Laplace mixture is the strongest method without a level at high
contamination, at $0.90$ or more up to $\varepsilon=0.35$ in D1, D3, D4, D6, D7 and D8, but
it fails with three and four groups already without outliers ($0.66$ and $0.68$) and with
clustered outliers.

\subsection{Covariate screen, other measures, further checks and cost}
\label{sec:measures}

\paragraph{The covariate screen.} It is what makes ESF robust to high-leverage points.
Without it (Supplement, Tables~S37 and~S38) the first step breaks down in D5 at
$\varepsilon=0.20$, as least squares does; outside D5 the two versions differ by at most $0.03$ up to
$\varepsilon=0.20$ in the first two parts, and from $\varepsilon=0.25$ on the screen gains up to $0.13$ and costs at
most $0.02$. The coefficient error (Supplement, Table~S19) tells the same story as the
accuracy: up to $\varepsilon=0.20$ the median error of ESF is at most $0.24$ in every design
D1 to D8, and it exceeds $1$ only where its accuracy has broken down.

\paragraph{The flagged fraction.} Where ESF works, its mean flagged fraction $\hat\alpha$
(Supplement, Table~S31) lies within about three percentage points of $\varepsilon$, and the
flagged set contains at least $89\%$ of the outliers and about $3\%$ of the clean units or
fewer (Supplement, Table~S16); where ESF breaks down it falls well below $\varepsilon$, because
the outliers are absorbed into a fitted line, which is the warning sign. After TCLUST-REG at
either conservative level, reweighting and the recovery step, the flagged fraction is as
close to $\varepsilon$ as that of ESF (mean absolute deviation over the $89$ cells with
$\varepsilon\le0.20$: $0.016$ and $0.017$ at the two levels, against $0.018$).

\paragraph{Further checks.} Supplement Section~S6 and Table~S39 report TCLUST-REG with equal
scales, with adaptive second-level trimming in the covariates and with a level chosen from its
monitoring path, the mixture of $t$ regressions \citep{yao2014robust}, robust linear grouping \citep{garciaescudero2009robust}, the incremental
reweighting of \citet{dotto2018reweighting}, two changes to our own constants, and the
mixtures with more starts (Tables~S21 and~S27). None gave a TCLUST-REG without a level or did
better than the pipelines above: the monitoring rules chose either a level below the
contamination or, nearly always, $0.40$, and the incremental reweighting fell below $0.85$ in
$22$ cells with $\varepsilon\le0.20$ after the recovery step, against four for the single
step.

\paragraph{Cost.} On one core ESF took a median of $0.08$ seconds per data set and about two
seconds with three covariates, most of it in the covariate screen; the times of the
competitors are in Supplement Table~S32.

\subsection{Sample size, groups, covariates and outliers}
\label{sec:further}

The third, descriptive part of the study ($2{,}300$ data sets, $50$ per cell) varied what the
first two held fixed: the sample size ($n=100$ and $n=1000$); the number of groups (four
parallel lines four noise scales apart, $n=400$, $m=16$); dependent covariates; moderate
outliers shifted by $4$ to $8$ noise scales, and outliers in a tight cloud; groups of $80\%$
and $20\%$ ($m=16$) and of $85\%$ and $15\%$ ($m=20$); and $t_3$ errors. Plan~5 added skewed
errors, groups of $70\%$ and $30\%$ with three covariates ($m=16$), and an error scale growing
fourfold over the design range (Supplement, Table~S25 and Section~S6). The sample size,
dependent covariates and moderate outliers change little: ESF stays within $0.014$ of the best
competitor chosen afterwards, except that with moderate outliers in D3 at $\varepsilon=0.30$ it
trails TCLUST-REG at level $0.25$ by $0.04$. Small groups are what the recovery step is for:
ESF alone reaches only $0.82$ without outliers and $0.73$ at $\varepsilon=0.20$ with groups of
$80\%$ and $20\%$, and between $0.60$ and $0.73$ with groups of $85\%$ and $15\%$, because its
best-scoring fit puts two nearly parallel lines on the large group and flags most of the small
one (Section~\ref{sec:algorithm}). With the recovery step it reaches $0.88$ to $0.91$ up to
$\varepsilon=0.20$, within $0.035$ of the best competitor, while at $\varepsilon=0.30$ the
first step flags more than a quarter of the units, the step rarely applies, and the accuracy
stays at $0.75$. Heavy-tailed, skewed and heteroscedastic errors leave the accuracy within
$0.006$ of the best competitor in every cell, but the flagged fraction then counts the tails
or the spread of the errors as well as the outliers: it exceeds the contamination by $6.5\%$
of the units with $t_3$ errors and no outliers and by $1$ to $6$ percentage points in Plan~5. Clustered outliers are the hard
case: at $\varepsilon=0.20$ ESF alone fits the cloud with one of its two lines and falls to
$0.70$; the recovery step repairs $11$ of the $27$ fits that fail in this way and raises the
mean to $0.78$, against $0.90$ for TCLUST-REG at a level above the contamination, and at
$\varepsilon=0.30$ no method of Table~\ref{tab:mainhigh} exceeds $0.72$.

\subsection{The recovery step: confirmation, sensitivity and repeated runs}
\label{sec:gate2}

The recovery step was developed on data generated with other seeds and frozen before Plan~3,
which applied it to the stored fits of ESF alone on the $6{,}400$ data sets the study then
held. All three criteria were met: in the six small-group cells with $\varepsilon\le0.20$ the
step raised the mean accuracy by $0.07$ to $0.31$ (at least $0.05$ was required), in no cell
with $\varepsilon\le0.20$ did it lower it by more than $0.004$ ($0.01$ was allowed), and the
comparison of Plan~1 was passed (Supplement, Table~S36). On the $600$ new data sets of Plan~4,
with two small-group designs not tried before (Supplement, Table~S41), the step raised the
accuracy by $0.09$ to $0.28$ in all $12$ cells, but with $20\%$ of outliers in the two new
designs ESF was $0.08$ and $0.09$ below TCLUST-REG at level $0.25$ with reweighting and the
recovery step ($0.91$ and $0.97$); in the other ten cells it was within $0.024$ of it.

\paragraph{Sensitivity and repeated runs.} Moving each constant of the recovery step one step
down and one step up (Supplement, Tables~S23 and~S44) changed no cell mean by more than about
$0.03$, except the limit on the flagged fraction (Section~\ref{sec:constants}). The noise range
$R$, $1.1$ times the range of the response, is set by the most extreme units, and a larger range
makes the uniform component less likely, so that a band of outliers passes more easily as a
group. With $R$ doubled, as if the most extreme unit lay twice as far out, the mean accuracy of
ESF up to $20\%$ of outliers moved from $0.914$ to $0.912$, but clustered outliers at
$\varepsilon=0.20$ lost $0.09$; with $R$ five times larger the losses reached $0.14$ with uniform
noise at $\varepsilon=0.20$, and after TCLUST-REG at level $0.40$ up to $0.29$ beyond $25\%$
(Supplement, Table~S48). Two ranges tied to the fitted scales, $20$ median scales and the range
of the unflagged units plus $10$ median scales, remove this dependence but make the uniform
component more likely everywhere and lose small groups: the mean accuracy of ESF up to $20\%$
fell to $0.891$ and $0.903$, with $12$ and $9$ cells below $0.85$ against two. The range of the
paper is the best of the three on these designs, at the price of a dependence on the most
extreme units. ESF depends on
its seed: in five runs with different seeds on each data set, the accuracies differed by more
than $0.05$ on $3.5\%$ of the data sets with up to $20\%$ of outliers and on $13\%$ of all
data sets, most of them where ESF breaks down (Supplement, Section~S6); this variation is of
the size of the thresholds of the plans, so the paired comparisons resolve differences of
about $0.03$ and no finer.

\paragraph{The recovery step after a trimming method.} After TCLUST-REG at level $0.40$ with
reweighting, the recovery step repairs the failures of the level with unequal groups up to
$\varepsilon=0.20$ (Section~\ref{sec:low}; Supplement, Table~S40) but not with three or four
groups, and it harms it materially once in the study (elsewhere by at most $0.016$):
with uniform noise at $30\%$ and $35\%$ it lowers the accuracy from $0.91$ to $0.87$ and from
$0.90$ to $0.72$, on $47$ of the $100$ data sets at $35\%$ by more than $0.05$, because a dense
and narrow band of uniform noise passes the peak test (Section~\ref{sec:recovery-theory}).
Moving each constant of the step did not prevent it (Supplement, Section~S5).
After the trimmed cluster-weighted model with reweighting (Supplement, Table~S46) the step did
the same: up to $\varepsilon=0.20$ it reduced the cells below $0.85$ from $39$ to $12$ at level
$0.25$ and from $66$ to $20$ at level $0.40$, and it lowered a cell mean by more than $0.02$
only with uniform noise at $30\%$ and $35\%$ (by up to $0.06$) and once with two parallel
lines at $30\%$ ($0.025$).

\section{Three data sets}
\label{sec:realdata}

\subsection{The fishery data}
\label{sec:fishery}

The fishery data \citep{riani2008fitting}, distributed with FSDA, record $677$ monthly import
flows of one fishery product into the European Union from a third country: the quantity and
the value of each flow. They are the standard example for TCLUST-REG and the forward search
\citep{perrotta2009new,cerioli2014robust}: flows priced differently form lines through the
origin, and a number of flows fit none of them. We work on the logarithmic scale, where value
$=$ price $\times$ quantity becomes a line of slope one whose intercept is the log price, the
spread of the errors no longer grows with the quantity, and the concentration of small flows
near the origin, which \citet{cerioli2014robust} handle by thinning, disappears. Twenty-six
flows with zero quantity or value have no logarithm and are left out, which leaves $651$. The
published analyses \citep{perrotta2009new,cerioli2014robust,torti2019assessing} keep them and
work on the raw scale, where the heteroscedasticity and the dense region near the origin are
part of the problem; the analysis below therefore answers a different question and is not a
comparison with their results. We fit $K=2$ groups with the constants of Section~\ref{sec:constants} and
$m=8$.

Supplement Table~S42 and Figure~S1 show the results. ESF alone finds two lines
of slope close to one, with intercepts $2.57$ and $2.85$, that is prices of about $13$ and
$17$, and flags $11\%$ of the flows, about $80\%$ of them priced below both lines. The recovery
step finds that these cheap flows lie close to a line of their own: for $19$ of $20$ seeds it
replaces the line at $2.57$ by a line with intercept $2.16$, a price of about $9$, keeps one
line through the two dearer price levels (intercept $2.78$), and flags $3\%$ of the flows.

TCLUST-REG gives the same two answers, but which one depends on the trimming level: at levels
up to $0.08$ it treats the cheap flows as a group, with intercepts between $2.2$ and $2.3$, as
ESF does; from level $0.20$ on it separates the two dearer price levels, as ESF alone does,
and its estimates then vary from one random start to another (Supplement, Section~S5). With the long search, reweighting and the recovery step,
level $0.25$ returns the cheap group and the dearer line (intercepts $2.15$ to $2.16$ and
$2.78$, $3\%$ of the flows flagged) from all ten starts, and level $0.40$ from three of ten;
without the recovery step both levels separate the two dearer price levels.

The data suggest three price levels. With $K=3$ and $m=16$, ESF returns lines with intercepts
$1.90$ to $1.91$, $2.50$ to $2.54$ and $2.82$ to $2.83$, slopes between $0.94$ and $1.00$, and
$3\%$ of the flows flagged, for $12$ of $20$ seeds; the other seeds give other fits. The
best-of-five rule of Section~\ref{sec:constants} returned the cheap group and the dearer line
with $K=2$, and the three price levels with $K=3$, every time over twenty disjoint sets of
seeds (Supplement, Section~S5). The Bayesian information criterion of the same model
prefers $K=2$ (Section~\ref{sec:discussion}); whether the cheap flows are a price level or
anomalous declarations is a question for the subject-matter expert.

\subsection{The tone perception data}
\label{sec:tone}

The tone perception data \citep{deveaux1989mixtures}, distributed with the R package
\texttt{mixtools} \citep{benaglia2009mixtools}, record $150$ trials in which a musician tuned a
tone to the octave above a fundamental with stretched overtones; the tuned ratio lies either
near $2$ or near the stretch ratio, so the data form a nearly flat line and a line of slope
one. \citet{bai2012robust} added ten identical points at $(0,4)$, far
from both lines and outside the range of the covariate. We fit $K=2$ with the constants of
Section~\ref{sec:constants} and $m=8$, to the original data and with the ten points added
(Supplement, Table~S35). On the original data every method finds the two lines, except the
trimmed cluster-weighted model at level $0.25$, which returns other fits from two of its ten
starts. With the ten points added, $6\%$ of the enlarged sample, one of the two lines is lost
from every start by TCLUST-REG at levels $0.02$ and $0.05$, by the trimmed cluster-weighted
model at level $0.05$ and by the Laplace mixture; the trimmed cluster-weighted model at levels
$0.10$ and $0.25$ loses it from three and six of its ten starts, and the noise-component
mixture for $7$ of $20$ seeds. ESF returns the same two
lines, to within $0.01$, for every seed, as do CTLE, the contaminated normal mixture and
TCLUST-REG from level $0.10$ on, with either search; the recovery step found no group among the
flagged trials. The flagged fraction of ESF, $0.25$ to $0.27$ on the original data, is not a
count of gross outliers: a tight core of trials sets the robust scale of the rising line, so
$18$ to $20$ trials of that line are flagged by their residuals, and the covariate screen flags
$14$ more (Supplement, Section~S5); TCLUST-REG with reweighting flags $19\%$. The fitted
lines are not affected.

\subsection{Taxi fares from an airport}
\label{sec:taxi}

The third example has a known answer. The New York City Taxi and Limousine Commission publishes
every yellow-taxi trip \citep{tlc2024trip}; we took the $157{,}264$ trips of March 2024 that
started at John F.\ Kennedy airport and carry a rate code. Their fare, before tolls, surcharges
and tips, follows one of two tariffs: a flat fare of $70$ dollars to Manhattan (rate code 2
with a valid record, $53\%$ of the trips), or the meter, which charges by distance and, in slow
traffic, by time (rate code 1, $38\%$); the remaining $9.6\%$ use other tariffs or are invalid
records (Supplement, Section~S5) and are treated as contamination. We regressed the fare on the
distance and the duration, $d=3$, with $K=2$ and $m=10$; nothing was removed, and the rate
code gives the true group of every clean trip. Twenty random samples of $2{,}000$ trips were
fitted by every method of Section~\ref{sec:simulation} except sequential RANSAC, which needs
the noise scale. The flat-fare group lies exactly on its line, so more than half of the units
have residual zero and the scale of the scoring step, a median of squared residuals, is zero;
the guard mentioned in Section~\ref{sec:design} takes the median over the units that a
replicate does not fit exactly.

Supplement Table~S43 gives the results. ESF found both tariffs in all twenty samples: the flat
fare exactly, and a metered line with median intercept $2.63$ dollars, $3.04$ dollars per mile
and $0.33$ dollars per minute, against $2.40$, $3.06$ and $0.33$ for the fit to all metered
trips. Its accuracy of $0.978$ is the ceiling that the best methods share: TCLUST-REG and the
trimmed cluster-weighted model at levels $0.10$ to $0.40$, the bisquare, contaminated normal and
noise-component mixtures all reach $0.975$ to $0.979$, as does TCLUST-REG with the long search,
reweighting and the recovery step at both levels, with smaller coefficient errors ($0.10$ and
$0.14$ against $0.23$); the recovery step changed none of its fits. TLE at levels $0.10$ and
$0.15$ and CTLE stopped with an error on most samples, presumably because the flat-fare group
has no variance; the example shows that the methods hold on a group without noise, not that
one is better than another. ESF took a median of $31$ seconds
per sample, most of it in the minimum covariance determinant of the covariate screen, for which
we used a general-purpose implementation (Supplement, Table~S43). Its flagged fraction, $0.23$, is more than twice the contamination: the metered fares depend on the
time spent in slow traffic, not on the duration, and their residuals have heavy tails, so
$16\%$ of the clean trips are flagged, more than half of them by the covariate screen, which
reads the long, skewed trips as outlying covariates; $89\%$ of the contaminated trips are
flagged. TCLUST-REG with reweighting, which has no covariate screen, flags $0.14$. The
best-of-five rule of Section~\ref{sec:constants} failed here once, because with a group that
has no noise the likelihood rewards any line through tied points.

\section{Discussion}
\label{sec:discussion}

The recovery step asks whether the units a robust fit has discarded hide a group, by the
likelihood of Gaussian groups plus uniform noise, and it can follow any method that flags
units.
After TCLUST-REG at a generous level with reweighting it returned the groups of $15\%$ to $30\%$
that the level had trimmed away, and it reduced the cells with a mean accuracy below $0.85$, up
to $20\%$ of outliers, from $31$ of $89$ to four; the four that remain have three or four
groups, where the level discards part of several groups at once, and the step left them at
$0.83$ and $0.81$. After the trimmed cluster-weighted model it did the same.
No fixed level of TCLUST-REG between $0.25$ and $0.40$, with or without equal weights, was
the best choice on both sides of $20\%$ of outliers; the level $0.35$ came closest, and was
within $0.03$ of ESF in $83$ of the $89$ cells up to $20\%$ but lost $0.11$ to $0.13$ with four groups. ESF, the flagging procedure without a trimming level, is
the option when that knowledge is missing and the contamination is probably below a quarter.
On the three data sets ESF reached, without a level, the answers that TCLUST-REG gives at the
right level.

The step settles whether a poorly fitted set of units is contamination or a group, by the
likelihood of the model that the flagging rule already assumes, one candidate at a time and at
the given $K$; where the number of groups is in doubt, a scan over the number of
groups (Section~\ref{sec:recovery}) is the better tool. After a generous level the
step accepted narrow bands of uniform noise as a group at $30\%$ and $35\%$ of outliers, so it
should not follow such a level when a large fraction of uniform noise is possible.

The flagged fraction $\hat\alpha$ is the quantity a user of a level-free method reads first,
and it is a contamination estimate only under near-Gaussian errors. Proposition~\ref{thm:step}
says what it estimates then, the contamination plus about $1.2\%$ of clean units, and on the
simulated data with Gaussian errors it lay within about three percentage points of
$\varepsilon$ wherever ESF worked. On the tone data ESF flagged $0.25$ to $0.27$ of a benchmark
that is treated as clean, and on the taxi data $0.23$ against a contamination of $9.6\%$,
because the errors are far from Gaussian in both (Section~\ref{sec:realdata}); the fitted
lines were right in both cases. A flagged fraction well above any plausible contamination
therefore says that the errors are not Gaussian, and one well below it says that the fit has
broken down (Section~\ref{sec:constants}).

Set ESF against TCLUST-REG at level $0.30$ with reweighting and the recovery step. Up to
$20\%$ of outliers the two have the same mean accuracy over
the $89$ cells, $0.914$ against $0.915$, and the fixed level has no cell below $0.85$ where ESF
has two, at $0.78$ and $0.83$ (Table~\ref{tab:level}). ESF gives three things. It needs no
level, and the level $0.30$ is a choice that pays only because the contamination in those
cells is below it: beyond $25\%$ the same level falls below $0.85$ in $17$ of $34$ cells,
ESF in $13$. With four groups it is more accurate, $0.95$ against $0.88$
(Table~\ref{tab:mainlow}), because the trimming level discards part of several groups at once
and the estimated group weights make this worse; the level $0.25$ with equal weights removes
that loss ($0.96$) but not the loss beyond $25\%$. It costs three things: a subsample size $m$,
which encodes a bound on the smallest group proportion or, at $m=12$, gives up a little
accuracy (Section~\ref{sec:constants}); the cap of $n/3$ on the flagged set, which stops the
recovery step where the fit is in doubt and so limits the recovery of small groups to about
$20\%$ of outliers; and a dependence on the seed that the fit of TCLUST-REG with $1000$ starts does
not show, which the best-of-five rule reduces at five times the cost.

The choice among the methods of the study is as follows. If the contamination is known to be
below a fifth, TCLUST-REG at level $0.25$ with equal weights, reweighting and the recovery step
was the most accurate pipeline, with ESF within $0.03$ of it in $85$ of $89$ cells. If it may exceed a
quarter, use a level of $0.35$ or $0.40$ with reweighting, followed by the recovery step unless the
outliers may be spread uniformly. If nothing is known about it, ESF with the recovery step
needs no level and is accurate up to about a fifth, and a flagged fraction well below the
plausible contamination warns that it has broken down. With two groups and outliers in the response only, the Laplace
mixture was about as accurate as ESF up to $20\%$ of outliers and more robust beyond, but it
fails with three or four groups, with high-leverage points and with clustered outliers.

Five limits should be kept in mind. ESF breaks down when outliers are so frequent that clean
subsamples are rare in the first stages and a contaminated fit is taken for the reference; in
our designs this happened from $25\%$ of outliers with uniform background noise or
high-leverage points and from $30\%$ with three groups, while with two groups of equal size,
or with three covariates, it held up to $35\%$. Small groups are recovered only up to about
$20\%$ of outliers, and not on every data set at $20\%$ (Section~\ref{sec:gate2}). Outliers concentrated in a tight cloud
were fitted as part of a group from $20\%$ on; the recovery step repaired about two fifths of
these fits. If the error spread grows with the covariate, units with large covariate values
are flagged: with a spread growing fourfold over the design range the flagged fraction
exceeded the contamination by up to $6$ points (Plan~5), and in raw trade data the spread
grows much faster, so a transformation that stabilises it, such as the logarithm of
Section~\ref{sec:realdata}, should precede the fit. The candidate search of the recovery step, a line through $d$ random units among $500$
draws, is itself unstable when the group to be recovered is a small part of the flagged set:
on the fishery data, after TCLUST-REG at level $0.40$ with reweighting, the step found the
cheap group from three of ten starts (Section~\ref{sec:fishery}), so the step can miss a
group that it would accept. Its noise range is set by the most extreme units, so that
outliers far out make bands of noise easier to accept as groups (Section~\ref{sec:gate2}).
A result for the whole procedure would need to control which units are flagged, and we do
not have one (Section~\ref{sec:theory}).

The number of groups $K$ was taken as known. The Bayesian information criterion of the
Gaussian-lines-plus-noise model over several values of $K$ gives a way to choose it: in a
descriptive check (Supplement, Table~S18) it chose the true $K$ for at least $90\%$ of the
data sets with two or four groups and up to $20\%$ of outliers, but for only $50\%$ to $80\%$
with the three overlapping groups of D2, and it counted a tight cloud of outliers as a group;
on the fishery data it prefers $K=2$ to $K=1$ and $K=3$.

\backmatter
\section*{Declarations}
\bmhead{Funding} The author received no specific funding for this work.

\bmhead{Competing interests} The author declares no competing interests.

\bmhead{Ethics approval} Not applicable.

\bmhead{Data availability} The fishery data are distributed with the FSDA toolbox
\citep{riani2012fsda}, and the tone perception data with the R package \texttt{mixtools}
\citep{benaglia2009mixtools}. The taxi data are public trip records of the New York City Taxi
and Limousine Commission \citep{tlc2024trip}; the extract used and the raw simulation output are
in the repository named under Code availability and in its Zenodo archive.

\bmhead{Code availability} The code, the scripts that produce every table and figure, and the
analysis plans with their SHA-256 digests are at \url{https://github.com/sorujov/eesp}
(``eesp'' is the name of an earlier version of the method), archived at Zenodo
(concept DOI \url{https://doi.org/10.5281/zenodo.22807276}, which resolves to the latest version).

\bmhead{Use of large language models} During the preparation of this work the author used
Claude (Anthropic), a large language model, to assist with writing and checking code, running
and managing the simulations, and drafting and revising the text. The author reviewed and
edited all such content and takes full responsibility for the article; the model is not an
author.

\begingroup\setlength{\bibsep}{1pt}\renewcommand{\normalsize}{\small}

\endgroup
\end{document}